# The Clustered Sparrow Algorithm

Cristian Dumitrescu
Independent mathematician, Kitchener, Canada.
cristiand43@gmail.com

**Abstract.** In this paper, we describe an algorithm for 3SAT. We use randomized rounding and stochastic local search in order to find a 2SAT expression that has the same satisfying solution as the original expression.

## 1 Introduction

The importance of efficiently solving NP – complete problems is due to the fact that this would imply that all the other problems belonging to the NP class can be efficiently solved in a constructive manner (the algorithms can be generated for all of them through polynomial - time reductions).

It is well known that 3SAT is NP-complete.

We will first present a well known randomized algorithm for 3SAT. This is Schöning's algorithm from 1991 (see [Schöning, 1991]). We also note that Papadimitriou also discussed a similar algorithm for 2SAT in 1991 (see [Papadimitriou, 1991]).

It is known that Papadimitriou's algorithm finds a solution in quadratic time with high probability for 2SAT. Only exponential bounds are known for Schöning's algorithm for 3SAT.

Stochastic local search algorithms played an important role, related to SAT solvers. This includes the random walk algorithm of Papadimitriou for 2SAT, and Schöning's extension of this algorithm for 3SAT.

We present *Schöning's* algorithm here because our new algorithm will have a similar structure, but it will work with different mathematical objects.
*Schöning's algorithm for 3SAT*

*Input: a 3SAT expression in* $n$ *variables.*

*Guess an initial truth assignment, uniformly at random.*

*Repeat* $3 \cdot n$ *times:*

> *If the expression is satisfied by the actual assignment, stop and accept.*
>
> *Let C be some clause not being satisfied by the actual assignment. Pick one of the three literals in the clause at random, and flip its truth value.*
>
> *Update.*

*Stop and reject, the expression is not satisfiable.*

Note that these algorithms involve random walks on the boundary of the truth assignment hypercube (the unit hypercube where each vertex represents a truth assignment).

## 2 The presentation of the algorithm

We consider the method of randomized rounding. For every Boolean variable x we keep track of a real valued r(x) ϵ [0, 1]. Initially these r(x) are all set to $\frac{1}{2}$ or to a random value between 0 and 1. At the end of the algorithm these values should be close to 0 or 1 and indicate a potential satisfying assignment. In each round of the algorithm any variable x is assigned the truth value 1 with probability r(x) and the truth value 0 with probability 1-r(x). So we have: $p(x = 1) = p(\neg x = 0) = r(x)$, $p(x = 0) = p(\neg x=1) = 1 - r(x)$. For a 3CNF with n variables , at each stage of the algorithm we have a set of n values, the set of r - values.

We consider a 3CNF expression. We consider a 3-clause $(x \vee y \vee z)$ , where the Boolean variables (or literals) x, y, z, have associated r- values r(x), r(y), r(z). So we have $p(x=1) = r(x)$, $p(y=1) = r(y)$, $p(z=1) = r(z)$, $p(x=0) = 1\text{-}r(x)$, $p(y=0) = 1\text{-}r(y)$, $p(z=0) = 1\text{-}r(z)$. We want to calculate the probability that $(x \vee y \vee z) = 1$ under these assumptions.

In general for 3 events A, B, C, we know that:

$$P(A \vee B) = p(A) + p(B) - p(A) \cdot p(B)$$

$P(A \vee C) = p(A) + p(C) - p(A) \cdot p(C)$

$P(B \vee C) = p(B) + p(C) - p(B) \cdot p(C)$

$p(A \vee B \vee C) = p(A) + p(B) + p(C) - p(A) \cdot p(B) - p(A) \cdot P(C) - p(B) \cdot p(C) + p(A) \cdot p(B) \cdot p(C)$.

In our case the event A is $(x = 1)$, event B is $(y = 1)$, event C is $(z = 1)$, so we have:

$p(x \vee y \vee z = 1) = p(x=1) + p(y=1) + p(z=1) - p(x=1) \cdot p(y=1) - p(x=1) \cdot p(z=1) - p(y=1) \cdot p(z=1) + p(x=1) \cdot p(y=1) \cdot p(z=1) = r(x) + r(y) + r(z) - r(x) \cdot r(y) - r(x) \cdot r(z) - r(y) \cdot r(z) + r(x) \cdot r(y) \cdot r(z)$.

$p(x \vee y \vee z = 1) = r(y) + r(z) - r(y) \cdot r(z) + r(x) \cdot (1 - r(y) - r(z) + r(y) \cdot r(z)) = p(y \vee z = 1) + r(x) \cdot (1 - p(y \vee z = 1)) = p(y \vee z = 1) + r(x) \cdot p(y \vee z = 0)$.

Analogously we have:

$p(x \vee y \vee z = 1) = p(y \vee z = 1) + r(x) \cdot p(y \vee z = 0) = p(x \vee z = 1) + r(y) \cdot p(x \vee z = 0) = p(x \vee y = 1) + r(z) \cdot p(x \vee y = 0)$.

So in general , if a variable x appears in a positive literal in a 3-clause and we want to increase the probability that the 3-clause $(x \vee y \vee z)$ has truth value 1 then we have to increase the r – value $r(x)$. On the other hand we have for example:

$p(\neg x \vee y \vee z = 1) = p(y \vee z = 1) + (1 - r(x)) \cdot p(y \vee z = 0)$

That's because $r(\neg x) = 1 - r(x)$.

So in general , if a variable x appears negated in a 3-clause and we want to increase the probability that the 3-clause $(x \vee y \vee z)$ has truth value 1 then we have to decrease the r – value $r(x)$.

For each 3-clause $(x \vee y \vee z)$) we consider the three 2-clauses (with two literals) $(x \vee y)$ , $(x \vee z)$ , $(y \vee z)$ that are associated with the initial clause with three literals. From each 3-clause we select one of the 2-clauses above, depending on the corresponding r – values. We define the r – value of a negated variable as $r(\neg x) = 1 - r(x)$, so we can talk about the r -value of a literal . In general if x, y, z are literals in a 3-clause and we have $r(x) > r(y) > r(z)$, then the 2-clause associated to the 3-clause above will be chosen to be $(x \vee y)$. If two literals have equal r - values, then we select one at random.

By an r – configuration we mean the set of r - values associated to the Boolean variables of our 3CNF expression. From each 3-clause we select a 2-clause as described above, with two literals that have the largest r – values. Thus we form a 2CNF expression associated to our 3CNF expression. So each r – configuration will generate a 2CNF expression associated to the initial 3CNF configuration.

Under a fixed satisfying truth assignment a satisfied 3-clause can have one, two, or all three literals taking the truth value 1. An associated 2 – clause is called a positive 2-clause if it has truth value 1 under a satisfying truth assignment. An associated 2 – clause is called a null 2-clause if it has truth value 0 under a satisfying truth assignment. At later stages of the algorithm we are trying to generate an associated 2CNF expression composed only of positive 2-clauses.

Each round of the algorithm starts with such a randomized rounding process. Each variable is assigned the truth value 1 or 0 following the probability distribution given by the corresponding r – values. Then we check whether the generated truth assignment is a satisfying assignment. We also check whether the current associated 2CNF expression admits a satisfying assignment. For this we need the routine 2SATSolver, which will find a satisfying assignment in at most quadratic time. We note that if the associated 2SAT expression admits a satisfying assignment , then this truth assignment will also satisfy the original 3SAT expression.

The other routine that we use is AdjustRValues. Under the current r-configuration which generated the current truth assignment we will have a set of unsatisfied 3-clauses. For all the variables x occurring in positive literals in these unsatisfied 3-clauses, the new r - values will be pushed close to 1. We use the relations $r(x)_{new} = 1 - \varepsilon - (1-r(x)_{old})^3$ . For each variable y occurring negated in these unsatisfied 3-clauses the new r - values will be pushed closer to 0. We use the relations $r(y)_{new} = r(y)^3_{old} + \varepsilon$, where $\varepsilon = 0.0001$ , for example. Other update formulas can be considered. The small quantity ε is used in order to make sure we don't hit 0 or 1 after only a few iterations due to the finite approximation level we work with.

*Algorithm for 3SAT*

*Input: a 3SAT expression in* $n$ *variables.*

*Uniformly at random generate an initial set of r – values.*

*Construct the associated 2SAT expression, by choosing a 2-clause from each 3-clause. The two literals chosen to form the selected 2-clause have the highest r – values in each 3-clause.*

*Repeat* $3 \cdot n^2$ *times:*

> *Consider the truth assignment generated by the current r – configuration.*
>
> *Call 2SATSolver on the associated 2SAT expression.*
>
> *If the expression is satisfied by the current truth assignment, or 2SATSolver finds a satisfying assignment, stop and accept.*
>
> *Consider all 3-clauses not being satisfied by the current truth assignment .*
>
> *Call AdjustRValues. Basically for all variables appearing in unsatisfied 3-clauses, increase the corresponding r – value (if the variable appears in a positive literal), or decrease the corresponding r – value (if the variable appears negated in the unsatisfied 3-clause).*
>
> *Call Update2SATExpression. Basically after AdjustRValues changes the r – configuration, update the 2SAT expression by choosing the 2-clauses from the corresponding 3-clauses such that the composing literals of the selected 2-clause have the highest r – values.*

*Stop and reject, the expression is not satisfiable.*

## 3 Comments.

We could consider a few versions of this algorithm. We could start with a set of r – values where r(x) = $\frac{1}{2}$ for all variables. We prefer to start with random values between 0 and 1. We try to avoid getting stuck in local maxima, so multiple restarts should improve our chances that we start close to a global maximum. Improved version of this algorithm will be discussed later.

Cristian Dumitrescu,
Kitchener, Ontario, Canada.

Email: cristiand43@gmail.com